# High-precision and low-latency widefield diamond quantum sensing with neuromorphic vision sensors

Zhiyuan Du[1]†, Madhav Gupta[1]†, Feng Xu[1]†, Kai Zhang[1,2], Jiahua Zhang[1], Yan Zhou[2], Yiyao Liu[3], Zhenyu Wang[3,4], Jörg Wrachtrup[5,6], Ngai Wong[1]*, Can Li[1]*, Zhiqin Chu,[1,7,8]*

[1]Department of Electrical and Electronic Engineering, The University of Hong Kong, Hong Kong, China

[2]School of Science and Engineering, The Chinese University of Hong Kong, Shenzhen, China

[3]Guangdong Provincial Key Laboratory of Quantum Engineering and Quantum Materials, School of Physics and Telecommunication Engineering, South China Normal University, Guangzhou 510006, China

[4]Frontier Research Institute for Physics, South China Normal University, Guangzhou 510006, China

[5]3rd Institute of Physics, Research Center SCoPE and IQST, University of Stuttgart, 70569, Stuttgart, Germany

[6]Max Planck Institute for Solid State Research, Stuttgart, Germany

[7]School of Biomedical Sciences, The University of Hong Kong, Hong Kong, China

[8]Advanced Biomedical Instrumentation Centre, Hong Kong Science Park, Shatin, New Territories, Hong Kong

*Corresponding authors. Email: nwong@eee.hku.hk; canl@hku.hk; zqchu@eee.hku.hk;

†These authors have contributed equally to this work.

**During the past decade, interest has grown significantly in developing ultrasensitive widefield diamond magnetometry for various applications, such as inspecting integrated circuits, measuring neuronal action potentials, and investigating vortex dynamics in superconductors. Despite attempts to improve the adoption of conventional frame-based sensors, including the lock-in technique, avalanche photodiode arrays, and the streaking imaging approach, achieving high temporal resolution and sensitivity simultaneously remains a key challenge. This is largely due to the transfer and processing of massive amounts of sensor data to capture the widefield fluorescence intensity changes of spin defects in diamonds. In this study, we adopt a neuromorphic vision sensor to address this issue. This sensor pre-processes the detected signals in optically detected magnetic resonance (ODMR) measurements for quantum sensing, employing a working principle that closely resembles the operation of the human vision system. By encoding the changes of light intensity into spikes, this approach results in a vast dynamic range, high temporal resolution, and exceptional signal-to-background ratio. After a thorough evaluation of theoretical feasibility, our experiment with an off-the-shelf event camera demonstrated a 13× improvement in temporal resolution with comparable precision of detecting ODMR resonance frequencies compared with the state-of-the-art highly specialized frame-based approach. A specialized camera system with the same mechanism has the potential to enhance these benefits further. This performance improvement is primarily attributable to orders of magnitude smaller data volumes and, thus, reduced latency. We further showcase the deployment of this technology in monitoring dynamically modulated laser heating of gold nanoparticles coated on a diamond surface, a recognizably difficult task using existing approaches. The current development provides new**

insights for high-precision and low-latency widefield quantum sensing, with possibilities for integration with emerging memory devices for more efficient event-based data processing.

**Introduction**

Solid-state quantum sensors have enabled new ways to detect magnetic, electric fields or temperature with extreme sensitivity that approaches the quantum limit [1, 2]. One of the most promising platforms so far has been the Nitrogen Vacancy (NV) center, an optically addressable defect in diamond, due to its exceptional electronic spin properties at room temperatures The electron spin state can be experimentally detected by the optically detected magnetic resonance technique, which involves sweeping the microwave (MW) frequency while recording the corresponding fluorescence intensities as a function of time[2]. Customized methods such as confocal-[3-7] and widefield-based[8-11] fluorescence microscope have served as gold standards for quantum sensing measurements. In particular, the widefield diamond quantum sensing approach allows for parallel readout of spatially resolved NV fluorescence, offering enormous potential in diverse fields[12-15]. However, this method generates a massive amount of data in the form of image frames that needs to be transferred from the camera sensors[11, 16-19] for further processing. This data transfer can significantly limit the temporal resolution, which is typically no more than 100 fps[20-22] due to the use of frame-based image sensors. As a result, the potential for widefield magnetometry in dynamic measurements has only been exploited to a limited extent. Several studies have proposed different approaches to improve the temporal resolution in widefield quantum sensing, including down-sampling method (with potential artefacts introduced)[23-25], frequency multiplexing[26, 27] (with complicated implementation while limited speed-up), advanced sensing arrays with single-photon avalanche diodes (SPADs)[28] (with complex circuit integration needed), and in-pixel demodulation with lock-in cameras[29, 30] (with sacrificed sensing precision). However, the fundamental limitation still lies in the monitored fluorescence intensity changes with image frames associated with a vast amount of data, leading to unsatisfactory performance in widefield quantum sensing. To overcome this bottleneck, we propose using a neuromorphic vision camera[31-33] to pre-process fluorescence intensity data near the sensor device, which reduces the data transmitted for post-processing and significantly enhances the temporal resolution, enabling fast dynamic measurements.

Unlike traditional sensors that record the light intensity levels, neuromorphic vision sensors process the light intensity change into 'spikes' similar to biological vision systems, leading to improved temporal resolution (~µs) and dynamic range (>120 dB)[31, 34, 35]. This approach is particularly effective in scenarios where image changes are infrequent, such as object tracking[36] and autonomous vehicles[37], as it eliminates redundant static background signals. Recently, this technique has gained attention in precision instruments measurements, such as emerging applications including fast-focusing in light microscope fast-focusing[38], dynamic magneto-optic Kerr effect (MOKE) microscopy[39], fast cell flow sorting[40], vibration measurement[41], fast-tracking of beads[42], and super-resolution imaging[43, 44]. Given that the fluorescence intensities encoded by microwave (MW) spatial-temporally vary only near the resonance frequency and therefore changes are rare, diamond quantum sensing is an ideal way of leveraging the benefits of this new approach.

This study, to the best of our knowledge, is the first to describe the application of the neuromorphic vision sensor to perform wide-field diamond quantum sensing. Specifically, we develop a custom and efficient protocol to process event-type quantum sensing data, which enables the reconstruction of derivative ODMR spectrum. Our experimental results demonstrate that this new approach takes far less time than conventional frame-based approaches (140ms vs 1.82s), while achieving similar precision (0.034MHz vs 0.031Mhz) in detecting the ODMR resonance frequency over a field of view (FOV) of 18*18 µm². We showcase its potential in monitoring sub-second scale laser heating of diamond surface coated with gold nanoparticles, which was previously impossible with conventional approaches. Temperature monitoring with 0.28s temporal resolution and 0.5K temperature precision is demonstrated in our experiment. We anticipate that our successful demonstration of the proposed method will revolutionize widefield quantum sensing, significantly improving performance at an affordable cost. Our study also paves the way for the development of intelligent quantum sensors with more advanced in-sensor processing capabilities[45-47], and brings closer the realization of near-sensor processing with emerging memory-based electronic synapse devices[48-50]. These advances hold great promises for further enhancing the performance of widefield quantum sensing, leading to new opportunities in scientific research and practical applications.

**Results**
**Neuromorphic widefield quantum sensing concept**

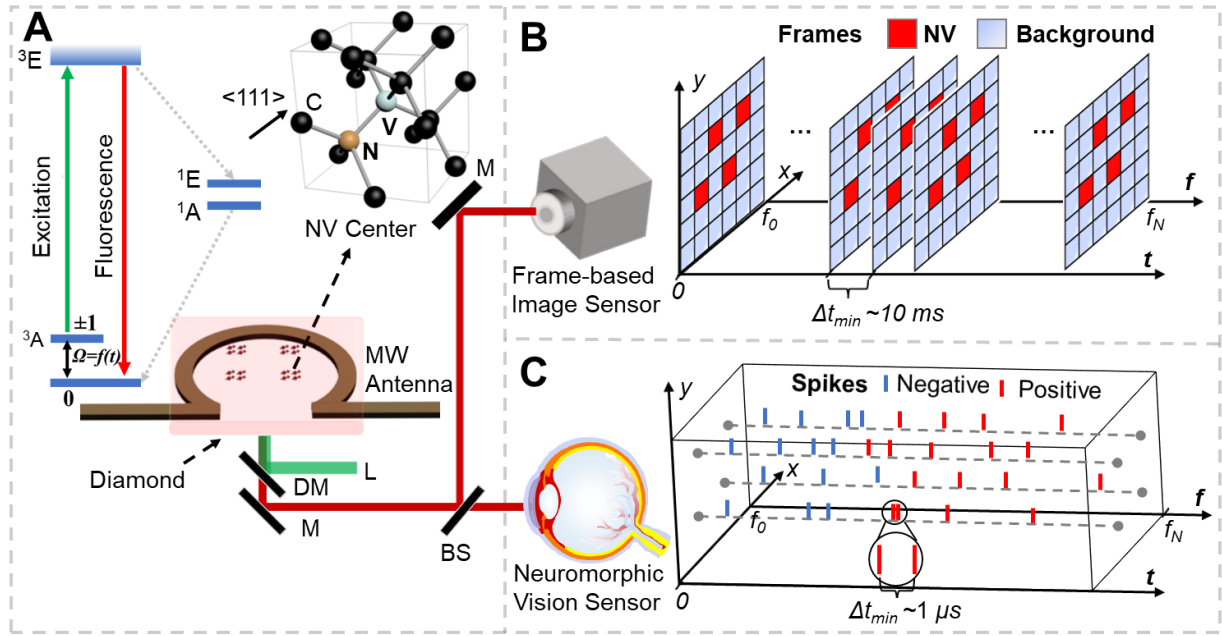

**Fig. 1. Concept, design and implementation of widefield quantum sensing.** (**A**) Overview of NV-based widefield quantum sensing: energy level diagram and atomic structure of NV centers; and the experimental apparatus of widefield quantum diamond microscope. L: Laser; DM: Dichroic Mirror; BS: Beam Splitter; M: Mirror; (**B**) A schematic showing the working principle of frame-based widefield quantum sensing, where a series of frames are output from a frame-based sensor recording both fluorescence intensity and background signals. (**C**) A schematic showing the working principle of proposed neuromorphic widefield quantum sensing, where the fluorescence changes are converted into sparse spikes through a neuromorphic vision sensor.

Diamond quantum sensing is facilitated by the NV center, which consists of a nitrogen atom and a nearby vacancy center hosted in diamond lattice. Due to the unique transition between triplets ground and excited states[2], the spin states of NV centers could be readout through the emitted red fluorescence excited with a green laser (Fig. 1A). The hallmark of quantum sensing based on NV centers in diamond is to perform the so-called optically detected magnetic resonance measurements, i.e., the monitored NV fluorescence changing with temporally encoded MW frequency. Specifically, the widefield ODMR measurement records the spatiotemporal NV fluorescence intensity changes in parallel, via a conventional frame-based sensor which normally operates at a limited framerate. With swept MW frequency, all pixels in the camera sensor synchronously record both regions of interest (i.e., NV fluorescence) as well as the background fluorescence, generating a series of frames with fixed time interval (Fig. 1B). This inflexible process produces highly redundant data (e.g., of the order of ~ 10 MB) for transmission and further process, causing a significant latency (e.g. of the order of ~10ms per frame). This makes it difficult to apply widefield diamond quantum sensing in many dynamic processes such as mapping the action potential of a single neuron[51].

The proposed widefield quantum sensing approach using a neuromorphic vision sensor aims to address the challenge described above. Instead of simply recording the fluorescence intensities from the frame-based camera, this method pre-processes data near the sensor. During widefield ODMR measurement, we observe that the fluorescence intensity only changes in the regions of interest and near the resonance frequency, while the majority of data changes only slightly. As a result, we adopt a neuromorphic event camera that converts the light intensity changes into sparse "change events" or spikes. (Fig. 1C). This resembles the working principle of photoreceptors in the human retina, which responds only to light intensity changes and converts them into spikes for transmission and processing in neural systems[32, 37]. For example, the working mode in our optical nerve has resulted in only about 20 Mb/s of data transmission to the visual cortex, while a rate of 20 Gb/s is required for the frame-based working mode in conventional digital cameras to match the same spatial-temporal resolution. The compressed data transmission thus results in significantly reduced latency and a high energy efficiency[32, 52]. Likewise, we use a neuromorphic sensor to measure the fluorescence change in parallel, from which a spike is generated only when the temporal fluorescence change surpasses a predefined threshold level. Because the fluorescence only changes significantly near the resonance, the general spikes are inherently sparse. Moreover, the spikes are only generated for the region of interest (ROI) where there exist intensity changes like NV centers modulated (temporally encoded) by MW frequency, further reducing the data transmission and improving the performance in widefield diamond quantum sensing.

**Event-based ODMR**

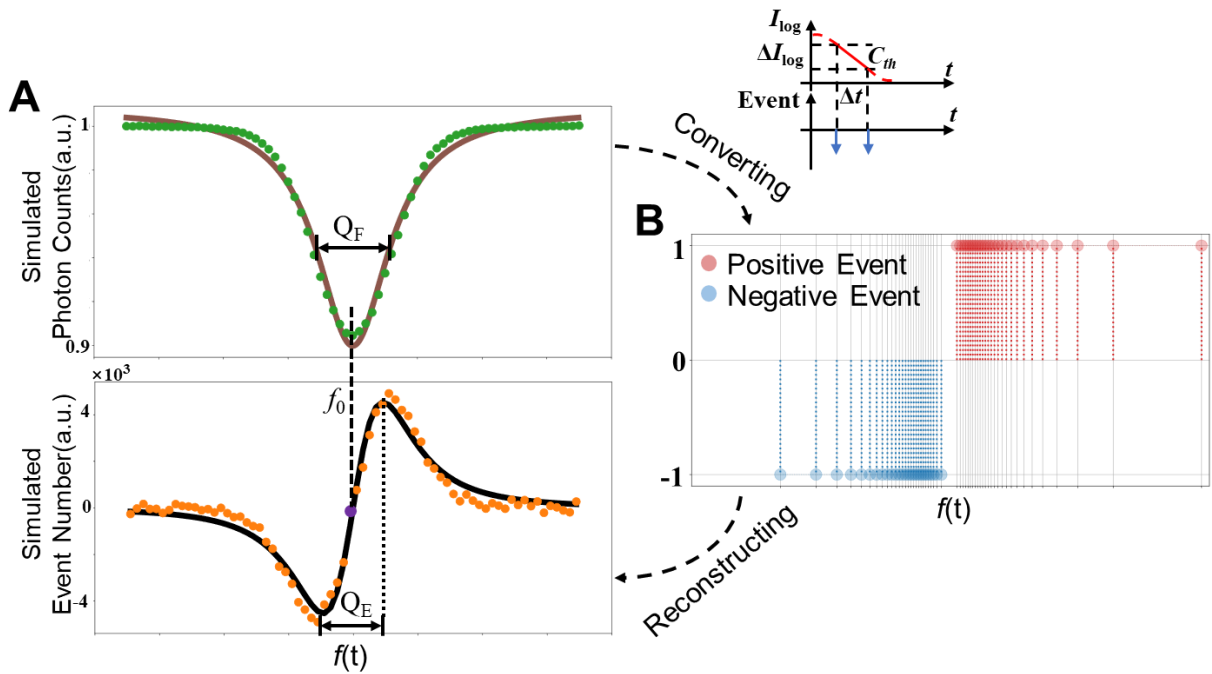

**Fig. 2: Theoretical background.** (**A**) Simulated ODMR spectrum using conventional frame-based sensor (green dots) and event-based sensor (orange dots), with quality described via $Q_F$ (upper panel) and $Q_E$ (lower panel), respectively. QF: full width at half maximum, QE: the frequency difference between two inflection points. The resonance frequency can be extracted by fitting the data with Lorentzian function (brown solid curve) and its derivative (black solid curve), respectively. (**B**) Cartoon showing the conversion from the frame-based ODMR spectrum into event-based one through processing the recorded time trace of computed raw events.

To demonstrate the feasibility of our idea, we performed Monte-Carlo simulations using a model with stochastic measurements as detailed in Methods. As shown in the upper part of Fig. 2A, our numerical simulation reproduced a light intensity that can be well-fitted with a Lorentz function that is consistent with previous experiments[3, 53]. For the proposed event-based approach, the simulated signal exhibits the shape of the derivative of a Lorentz function (lower part of Fig. 2A).

In fact, the relationship between the original and event-based ODMR can be well understood and mathematically derived: First, the light intensity is converted to a series of events in our simulation, based on the working process of the proposed neuromorphic sensor[31]. In this regard, each sensing pixel responds to light intensity changes independently and produces a positive event when the light intensity increase surpasses a predefined threshold $C_{th}$, and a negative event for a decrease (Fig. 2B). Therefore, if the threshold value is much smaller than the intensity change, the time interval $\Delta t$ between two events can approximately describe the derivative of the original spectrum at that point:

$$\Delta t = \Delta I_{\log} / \left(\frac{\Delta I_{\log}}{\Delta t}\right) = \frac{1}{I'_{\log}(t)} \times \Delta I_{\log} = \frac{1}{I'_{log}(t)} \times C_{th} \quad (1)$$

Equation (1) clearly shows that the fluorescence intensity gradient is encoded as the density of events, as $\Delta t$ is inversely proportional to the derivative of intensity, as illustrated in Fig.2B. To recover the spectrum gradient, we calculate the event density $\lambda_s(t)$ by counting the number of simulated events within a certain time range $T$:

$$\lambda_s(t) = \frac{T}{\Delta t} = \frac{T}{C_{th}} \times I'_{log}(t) \quad (2)$$

Finally, the derivative Lorentzian spectrum $I'_{log}(f)$ is reconstructed from the $I'_{log}(t)$ using an established relationship between time $t$ and microwave frequency $f$. This is clearly verified by our simulation results (shown in the lower panel in Fig.2A), where the discrete points of summed events can be fitted with the derivative Lorentzian function. The resonance frequency $f_0$ can also be determined at the point where the derivative crosses zero value. To describe the quality of the spectrum, the quality metric $Q_F$ (for original Lorentzian spectrum) and $Q_E$ (for derivative Lorentzian spectrum) can also be calculated, where $Q_F$ is defined as the full width at half maximum (FWHM), while $Q_E$ is defined as the frequency difference of two inflection points of the spectrum. Therefore, our method provides a new route to represent ODMR by the post-processed events. Consequently, we refer to this new technique as event-based ODMR measurement. This form of measurement has guided our following experiments.

**Experimental demonstration of Event-based ODMR measurement**

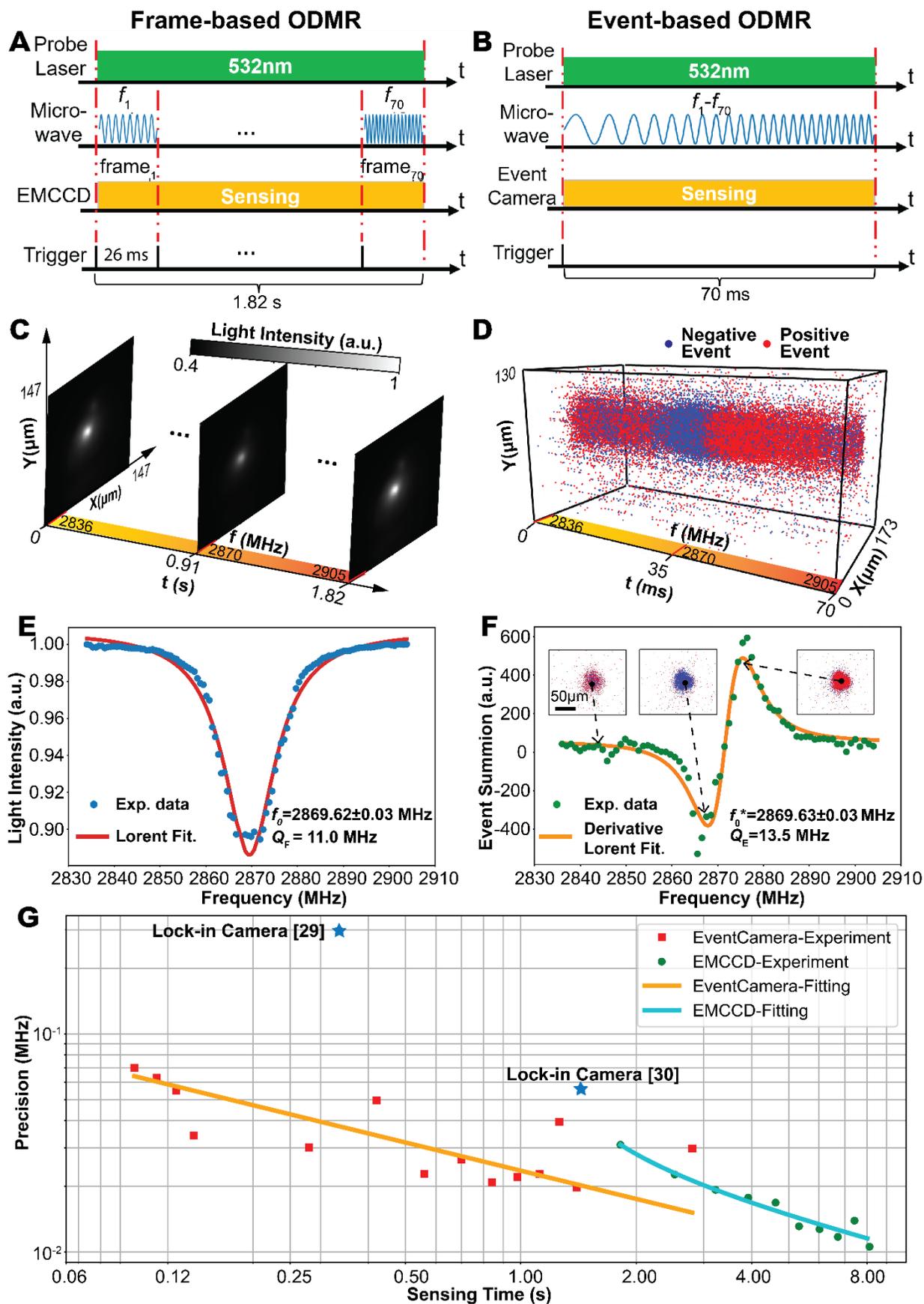

**Fig. 3**: **Experimental demonstration.** The measurement protocol, raw datasets and obtained ODMR spectrum (of the central point of ROI) using frame-based (**A**, **C**, **E**) and event-based sensor (**B**, **D**, **F**), respectively. The insert in (F) shows raw event frames (by accumulating events of 1ms

range) at three different frequency points. The spectra in (E) and (F) are fitted with the Lorentzian and its derivative functions, respectively, from which the resonance frequency $f_0$ is extracted ($f_0^*$ is the averaged result from forward and backward sweeping as discussed in Supplementary Section 1, error represents the standard deviation from 10 repeated measurements). **(G)** Comparison of precision σ and required sensing time τ for measurements using event camera (red squares), EMCCD (green circles) and lock-in camera (blue stars), respectively. The results of event- and EMCCD- based methods are fitted with $\sigma_{event} = 0.023 \frac{1}{\tau^{0.43}}$ (orange solid) and $\sigma_{EMCCD} = 0.028 \frac{1}{(\tau-\tau_o)^{0.48}}$ (cyan solid), respectively.

| Method | Sensing Time | Precision | Data Amount | Latency | SBR$_s$ | SBR$_t$ | Dynamic Range |
|---|---|---|---|---|---|---|---|
| Conventional Quantum Sensing (e.g., EMCCD[1]) | 1.82 s | 0.031 MHz | 35 MB | 26 ms | 64 | 1 | 96 dB |
| Neuromorphic Quantum Sensing (e.g., Event Camera[2]) | 0.14s* | 0.034 MHz | 363 KB | 220 μs | 194 | 10 | 120 dB |

1. Evolve 512 Delta Photometrics, price ~ $20000[54]; 2. EVK1-Gen 3.1 VGA Prophesee, price ~ $4000[55];

*. The calculation considers both forward and backward frequency sweep.

**Table 1. Comparison between frame-based and event-based ODMR**. The sensing time, precision and data amount are compared with their typical values obtained by experiment. SBR$_s$ and SBR$_t$ stand respectively for spatial and temporal signal-to-background ratio, defined in Supplementary Section 3.

We have successfully demonstrated our event-based ODMR measurement concept through experiments and systematically compared its performance with the conventional frame-based approach. As a benchmark for comparison, we used a highly specialized Electron Multiplying Charged Coupled Device (EMCCD), a typical frame-based camera used in traditional ODMR measurement. Due to its frame-based working mode, we swept the MW across 70 discrete frequency points to perform ODMR measurements with a framerate of 38.5 fps or 26 ms per frame (Fig. 3A). As a result, the overall measurement time was 1.82 s for one complete ODMR measurement. By contrast, the event camera is not limited by frame rate due to its unique working principle, so we swept the same frequency range in a linear chirp manner (continuously) with only 70 ms, as shown in Fig. 3B. (As illustrated in the section "Event-based ODMR measurement" in Methods, we repeat the sweep for 10 loops for one complete ODMR measurement to mitigate the influence of noise events). In fact, the time could be further improved with a trade-off with the sensing precision (discussed in detail in Fig.3G). Indeed, an extreme short time of 3.5 ms, with a degraded yet acceptable precision (0.11MHz) has been realized in our event-based experiment. Such a short time is unattainable by frame-based methods (Fig. 3G).

The superior performance of the proposed event-based ODMR measurement is attributable to its unique working principle, which could be explicitly seen from the raw data format. The frame-based widefield ODMR measurement generates a series of frames, representing a massive amount of data to transfer during the scanning of full ODMR spectrum across all pixels (Fig. 3C) in order to maintain the precision of the ODMR measurement. This results in a limited framerate of the camera and increased sensing time. By contrast, the proposed event-based wide-field ODMR measurement generates data in the form of sparse events, significantly reducing data transfer and enabling much faster sensing speeds. This feature is evident in our experimental data (Fig. 3D) during our event-based wide-field ODMR measurement. The data consists of a stream of spatial-temporal events in which the event density and polarity encode the information related to the fluorescence intensity changes. As expected, the detected event density is noticeably concentrated near two turning points, while it is sparse in the off-resonance region of the measured event-based ODMR spectrum (insert in Fig. 3F). A consistent demonstration for this phenomenon can be observed in the recorded video (Movie S1). This can also be verified by the time trace of accumulated number of events in one typical measurement (fig. S8, where the average event number from the central pixel is counted for every 1ms of the sweep). The event number remains minimal in the off-resonance frequency region, with only a few events generated due to noise. However, the number increases significantly near the in-resonance frequency (close to the turning points). These data are consistent with our assumption and further support the improved performance of our method.

The proposed event-based ODMR measurement employs a distinct measurement protocol, data representation, and processing method, but still achieves the same resonance frequency as frame-based ODMR measurements. For conventional frame-based ODMR measurements, the raw intensity recorded at a specific location, which varies with microwave frequency, can be fitted with a Lorentzian function to extract the resonance frequency ($f_0$=2869.62MHz) (Fig.3E). By contrast, the event-based measurement reconstructs the derivative Lorentzian spectrum from the density of events, i.e., the summation of event number over a defined period (Fig. 3F). By fitting with a derivative Lorentzian function, the resonance frequency can also be extracted, although it appears to exhibit a noticeable deviation ($f_0$=2872.23MHz) compared to the EMCCD result. This deviation is caused by the time delay between the event camera and MW source and also the large threshold of the event camera (detailed in Supplementary Section 1) and can be easily compensated for by performing another backward frequency sweep ($f_0$=2867.03MHz) and averaging the two results (the corrected resonance frequency $f_0^*$=2869.63MHz indicated in Fig. 3F).

The distinct measurement protocol also results in a different model to describe the trade-off relationship between the measurement precision $\sigma$ and the sensing time $\tau$. We conducted experiments to study this relationship with both frame-based ODMR measurement using an EMCCD camera and event-based ODMR measurement using an event camera. Details on precision calculation are provided in the section "Calculation of sensing precision and time" in Methods. From the result in Fig. 3G, it is evident that precision improves with longer sensing time for both methods. The trade-off for the frame-based approach can be explained with the shot-noise model[56] ($\sigma \propto \frac{1}{\tau_e^{0.5}}$), where $\tau_e$ is the total exposure time for one complete ODMR measurement. Our experiment shows that it follows: $\sigma_{EMCCD} = 0.028 \frac{1}{(\tau-\tau_o)^{0.48}}$, where $\tau - \tau_0$ is the exposure time $\tau_e$ in the short noise model, and $\tau_0$ is the overhead for data readout and transmission, so the result is closely aligned with the theoretical model. In our measurement, $\tau_0$ =1.12s (for readout of 70 frames), which is roughly consistent with the camera's fastest speed (67 fps)[54]. The fitting curve for our event-based method follows $\sigma_{event} = 0.023 \frac{1}{\tau^{0.43}}$. We attribute this root-inverse property to varying probabilities of event generation (see Supplementary Section 2), rather than the shot-noise limitation experienced by the EMCCD, given that event camera pixels detect photo-current rather than integrated photo-generated charges[55].

A comparison of the results of our event-based and frame-based methods clearly demonstrates that the event-based approach significantly reduces sensing time while maintaining a comparable level of precision. As illustrated in Fig. 3G, data points derived from the event-based method fall to the left of those obtained using the EMCCD-based frame approach. Moreover, the performance of lock-in camera-based works realizes either high precision but long sensing time[30] or a shorter sensing time with significantly degraded precision[29], but in general is not competitive with our results. Table 1 compares the key performance metrics of the event-based and frame-based methods, demonstrating that our data reduces the sensing time of the event-method by more than an order of magnitude (0.14s vs 1.82s) while maintaining a similar sensing precision (0.034MHz vs 0.031Mhz).

This time-saving mainly comes from the negligible data readout, as only a few hundred kilobytes of event data need to be transferred after the pre-processing near the sensor. The data transfer overhead is estimated to be ~1.7ms based on 1.6Gbps camera bandwidth[55], compared to 1.12s required for the frame-based method using an EMCCD (consumed mainly by the pixel-by-pixel analog-to-digital conversion (16bit) and readout).

The comparison also suggests that our approach has a much higher spatial and temporal signal-to-background ratio ($SBR_s$ and $SBR_t$[57]) than the conventional method (see Supplementary Section 3 for the calculation of these values based on raw events). The higher SBR values are attributable to the unique working mode of the neuromorphic method, which only responds to changed fluorescence. Since the light intensity from background pixels away from the ROI does not change, only rare events are produced by large noise. This helps to reduce data redundancy and also makes it easier to distinguish the ROI from the background area, which is highly desirable in nano-diamond related applications[58]. Finally, we note that the event camera can work in a wider dynamic range[31] and at a much lower cost than the EMCCD camera used for comparison. The performance of our method can, in principle, be further improved by adopting a high-figure-of-merit neuromorphic vision sensor[59].

**Widefield temperature dynamics measurement**
To showcase the potential application in monitoring highly dynamic processes, we experimentally demonstrated widefield NV-based quantum thermometry measurements. The NV-based quantum thermometry relies on the thermally induced ODMR spectrum shift, which has been recognized as an ultrasensitive platform for various scientific and industrial applications[13, 24, 60-62]. We began with a static measurement, where the power of the laser used to heat up the sample was fixed at specific values, ensuring the system is settled in an equilibrium state. The static temperature distribution was calculated from the ODMR resonance frequency shift. Our experiment revealed a linear relationship between measured temperature and heating laser power, as shown in Fig. 4B. The measurement precision was below 0.5 K for all measurements (lower panel in Fig. 4B), and an almost uniform temperature distribution (upper panel in Fig. 4B) was observed due to the high thermal conductivity of the bulk diamond sample[63].

Based on the static measurement, we further demonstrated the dynamic temperature monitoring, where the temperature is controlled via an electrically-rotated linear polarizer that tunes the heating laser power (red in Fig. 4A). By rotating the polarizer at a fixed speed $\omega$, the heating laser power irradiating the sample surface will be tuned in a continuous cosine square pattern (fig. S11A), indicating a similar pattern in temperature dynamics, i.e., $\Delta T = A_0 cos^2(\omega t + \varphi) + c$. With the event-based OMDR measurement, we achieved a temporal resolution of 0.28s, demonstrating an easy widefield temperature tracking. The periodic temperature change within the FOV is clearly observed in Fig. 4C (see also Movie S2). The temperature change is consistent with the heating laser power in terms of the cosine square fitting (the center pixel is shown in Fig. 4D while others in fig. S11B) and the Fourier transform (FT) (Fig. 4E). The fitting shows a bias from 0 K (reference zero point measured with 0 mW red laser in the static measurement) because of the accumulated heat during continuous laser tuning, which requires more time to be released to reach the equilibrium temperature. The extracted rotation speed from fitting (i.e. $\omega_{meas.}$= 0.728 rad/s) is very close to the pre-set value $\omega_{set}$ = 0.724 rad/s. Combining measurement from all pixels during the widefield measurement confirms spatially uniform temperature dynamics. By contrast, the results measured with the frame-based method using an EMCCD show aperiodic and smaller-range temperature change (lower panel in Fig. 4D) and also irregular FT (Fig. 4E), indicating a failure to track the temperature change due to the much longer sensing time (with temporal resolution of 1.82s). The sensing temporal resolution of our event-based method can be further improved to 0.14s at an expense of slightly reduced precision (fig. S11C).

Interestingly, we also discovered that the measured temperature amplitude using our method decreases with the increased rotation speed of the polarizer (summarized in Table 2 and fig. S12). This phenomenon is attributable to the long response time for the thermal dynamic property of the gold particles, instead of under-sampling, which has been verified by the transition measurement with temperature switched by an acousto-optic modulator (AOM). The periodic temperature switches are reproduced matching the protocol shown in fig. S13. The heating and cooling process of the first cycle is fitted with a first-order exponential response function, from which we extract a 0.71s rising and falling time. The first-order frequency response based on this response time aligns well with the previously mentioned temperature amplitude measured under different rotation speeds, a phenomenon that cannot be observed with the frame-based method using an EMCCD.

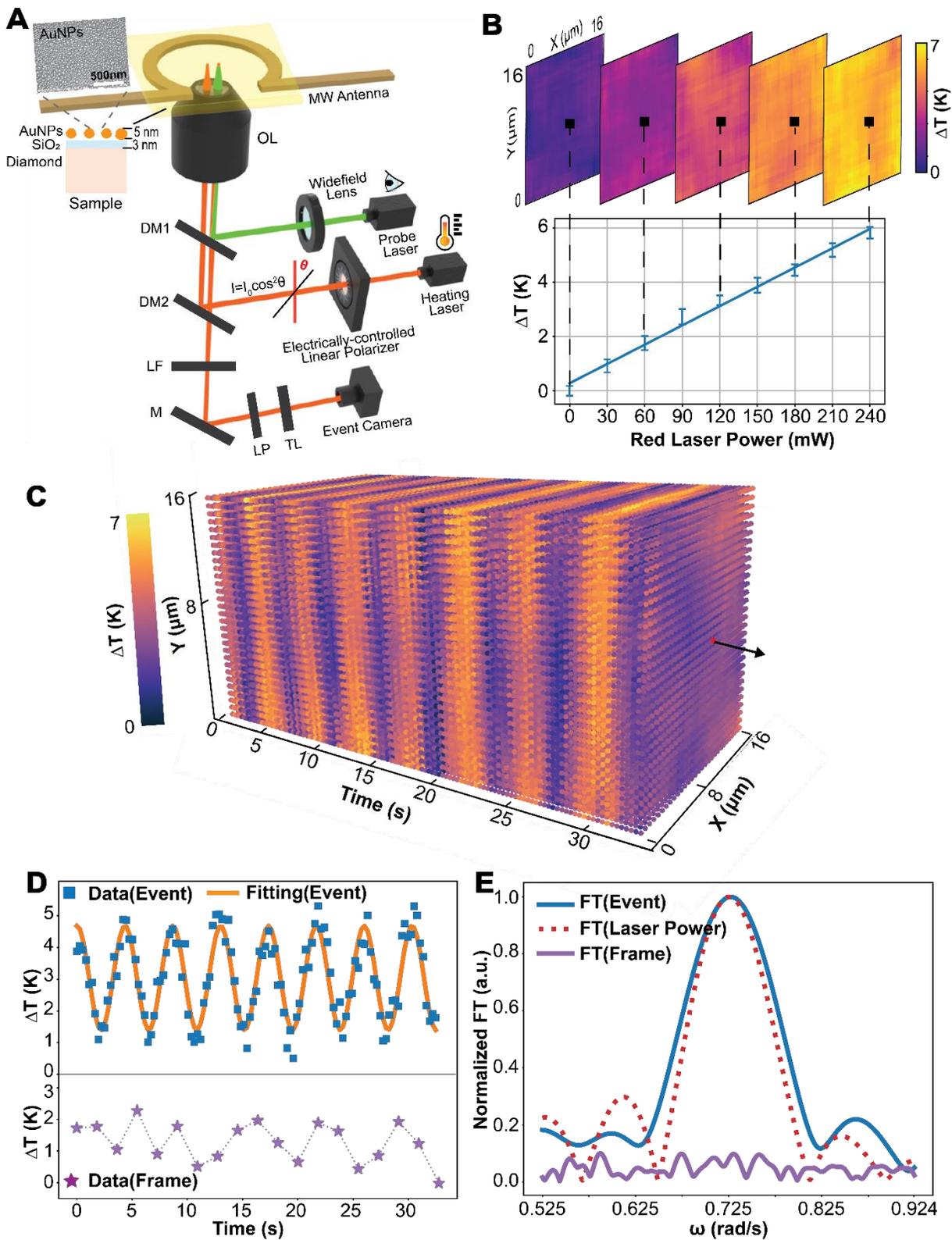

**Fig. 4. Widefield dynamic temperature measurements.** (**A**) Setup for dynamic temperature measurement. The main part of the system resembles that shown in fig. S1A, with an additional red laser serving as the heating source tuned by an electrically rotated linear polarizer. (**B**) Static measurement of temperature change vs red laser power for the central point of ROI. (**C**) The spatiotemporal temperature response of the sample measured with event-based method. (**D**) Cosine temperature change in the central point of ROI measured with event-based (blue squares) and frame-based (purple stars) ODMR. Only the event-based method tracks the true temperature change which can be fitted with the cosine square function. (**E**) Fourier transforms (FTs, magnitude) of the data in D. (the FT of frame-based results is scaled to 0-0.1 for a clear comparison). The FT of event-based results is consistent with that of laser power tuned by the same rotated polarizer.

Table 2. Measured rotation speed and temperature change range under different modulation speeds of the polarizer in comparison with the set values

| $\omega_{set}$ (rad/s) | $\omega_{meas.}$ (rad/s) | $A_0$ (K) |
|---|---|---|
| 0.207 | 0.206 | 5.56 |
| 0.414 | 0.415 | 4.61 |
| 0.724 | 0.728 | 3.34 |

**Discussion**

The essence of widefield quantum sensing is to detect changes in the number of photons across space and time, presenting a complex trade-off problem in both spatial and temporal domains. Our event-based working process holds the smart pre-process capability that detects sparse events adaptively in both space and time, thus matching well with the requirement of quantum sensing. Specifically, the neuromorphic pixels work independently and asynchronously, enabling the immediate readout of detected fluorescence change without waiting for the other pixels, which allows for an extremely high time resolution. Moreover, the event data that constitute the time-varying fluorescence spectrum have an adaptive time interval because events are generated only when the light change surpasses a threshold. This efficient process reduces redundant data and overcomes the limitations of frame rate in the frame-based approaches, enabling low-latency ODMR measurements.

It should be emphasized that our method has significant potential for further development in the future. In addition to its application in dynamic temperature measurement, it can be readily extended to magnetic field sensing, which has implications for the manipulation of magnetic skyrmions[12, 64, 65], spin-assisted super-resolution imaging[66-68], and detection of neuron action potential[13, 51], among other possibilities. Furthermore, neural network algorithms[69-72] could be used to map the raw events back to the original spectrum, as they preserve the derivative function relationship, or directly infer the observables such as temperature and magnetic field., potentially optimizing the precision further. Integration of electronic synapse devices[48-50] could also enable in-sensor or near-sensor algorithm execution[45-47], paving the way for the development of intelligent quantum sensors.

To summarize, we have demonstrated an event-based quantum sensing method that achieves both low-latency and high-accuracy ODMR measurement. A derivative Lorentzian spectrum can be reconstructed from raw events that are transferred from continuous fluorescence change through an event camera. By fitting the equation, the resonance frequency is extracted with comparable precision (0.034MHz vs 0.031MHz) but in a much shorter time (0.14s vs 1.82s) than the results obtained from frame-based ODMR using an EMCCD. The working principle also offers additional benefits, such as adaptive sampling, higher SBR and a wider dynamic range. Finally. our method is successfully demonstrated in tracking widefield dynamic temperature change with a 0.28s time resolution.

**Materials and Methods**

**Sample preparation.** The diamond sample was bought from Element Six (UK) Ltd. (SC Plate CVD 3.0x3.0x0.25mm <100> P2 145-500-0549). On top of the sample, a thin layer of gold nanoparticles was fabricated as follows: Gold nanoparticles were synthesized according to a published process[73]. Briefly, mixed solutions of NaOH (5 ml, 0.1 M) and ultrapure water (45 ml) were prepared. Tetrakis(hydroxymethyl)phosphonium chloride (THPC, 67.2 μmol in 1 ml water) was added to the above mix solutions. After 5 mins, HAuCl4 solution (2 ml of a 1% w/w solution in water, 59 µmol) was added under vigorous stirring. The seeding gold colloid solution was obtained. Single-crystalline bulk diamonds were sonicated in acetone and isopropanol for 5 min each and dried with nitrogen. The diamonds were cleaned and chemically activated by freshly prepared piranha solution (H2SO4/H2O2=7:3) at 90°C for 1 hour, rinsed thoroughly with ultrapure water and ethanol, and dried with nitrogen. 1,2-bis(triethoxysilyl)ethane (10 ul, BTSE), tetraethoxysilicicic acid (20ul, TEOS) and 3-aminopropyltriethoxysilane (20ul, APTES) were slowly added dropwise to a mixture of ethanol (2850 µl), ultrapure water (150 µl) and hydrochloric acid (10 µl), and hydrolyzed for 2 hours. After hydrolysis, the cleaned diamonds were placed into the above hydrolysis solution and deposited for 6 hours. After the reaction, the diamonds were cleaned with ethanol and dried with nitrogen. Mixed solutions of hydrochloric acid (10 µl) and ultrapure water (1 ml) were prepared. 1ml seeding gold colloid solution was added the above mix solutions. Diamond with surface amination was placed into the above good colloid solution and deposited overnight. After the reaction, the diamonds were cleaned with ultrapure water and dried with nitrogen and obtained diamond with gold film.

**Measurement Setup.** fig. S1A shows the setup we used for performing ODMR. We used a 532nm laser (MGL-III-532) to excite the diamond sample (prepared using the procedures mentioned above). After being expanded by a widefield lens, and reflected by a dichroic mirror (DM, cut-off wavelength 605nm), the laser illuminated the diamond sample through a microscope objective lens (OL, Olympus UplanSApo, 40x/0.95NA) with a 18um (FWHM) beam spot. The emitted fluorescence was then collected by the same objective lens. Microwaves (MW) were generated by a custom-built system shown in fig. S1C, where microwave signals from a RF signal generator (SynthNV PRO, with frequency $f_1$ fixed at 2835MHz) and an arbitrary waveform generator (AWG, Rigol DG5071, with frequency $f_2$ swept from 1 to 70MHz) were mixed through a RF mixer (Mini-Circuits ZEM-4300+) to yield the target frequency $f_1+f_2$. After further amplification using a microwave amplifier (ZHL-16W-43-s+), the mixed signal was fed on diamond through a waveguide for tuning the NVs' spin states. The tuned fluorescence was first filtered by the Long-pass filter (LP, cut-off wavelength 650nm), and then detected by an event camera (Prophesee, EVK1-Gen3.1 VGA) after passing through the tube lens (TL). For comparison, we built anther optical path in the same system for traditional quantum sensing using EMCCD (Photometrics, Evolve 512 Delta), which can be switched through a flip mirror (FM). Moreover, a series of pulses were generated through a digital pulse generator (Quantum Composer, 8210) to synchronize the MW frequency sweep and the cameras' measurement.

**Details of simulation.** In our simulation, we modeled the NV electron spin dynamics by including stochastic projective measurements during the protocols to take the continuous ODMR measurement process into account. Specifically, we used a weak measurement rate $\Gamma_M$ to describe the measurement speed such that the probability of no projective measurement occurring within a short time interval $\Delta t$ was $e^{-\Gamma_M \Delta t}$. Between successive projective measurements, the dynamics of the NV electron spins was driven by the NV Hamiltonian that includes the effect of microwave control. When a projective measurement occurred, it would prepare the NV electron to the $|m_s = 0\rangle$ spin state due to optical initialization and emit a photon with a probability $p_\text{photon} = |\langle 0|\Psi\rangle|^2$, where $|\Psi\rangle$ is the state just before the projective measurement. The intensity signal $I$ is proportional to the counted number of photons.

The probability distribution of the time $t_{2M}$ between two successive projective measurements is an exponential distribution. This allows us to randomly generate $t_{2M}$ by using a random variate $p_M$ drawn from the uniform distribution on the unit interval (0,1) with the relation $p_M = 1 - e^{-\Gamma_M t_{2M}}$. We set $\Gamma_M = 1.5$ MHz to match the experimental results. To reduce the simulation overhead, we assume that the evolution of each NV electron spin state $|\Psi\rangle$ between two successive projective measurements was driven by a simplified two-level Hamiltonian $H = \frac{\hbar}{2}(\Delta\sigma_z + \Omega\sigma_x)$, where $\Omega$ is the Rabi frequency of microwave control and $\Delta$ is the detuning between the microwave frequency and the energy splitting $\omega_{NV}$ of the NV $m_s = 0, -1$ spin states. $\sigma_z$ and $\sigma_x$ are Pauli operators with $\sigma_z = |0\rangle\langle 0| - |-1\rangle\langle -1|$. In the simulation, we considered 10000 NV centers where the values of $\omega_{NV}$ follow a normal distribution with mean value $\mu = 2\pi \times 2870$ MHz and standard deviation $\sigma = 2\pi \times 5.5$ MHz. The aforementioned steps were repeated until the continuous ODMR measurement time $T$ was reached.

For the original ODMR, we varied the microwave frequency $\omega_{MW}/(2\pi)$ discretely from 2836 MHz to 2905 MHz with a step size of 1 MHz. The measuring time for each frequency is $T_{frame} = 10$ ms. These values were used in the experiments. The light intensity corresponding to each frequency was obtained by summing up the photons generated by all NV centers within the time $T_{frame}$. The light intensity was normalized and displayed in the upper panel of Fig.2A.

For the simulation of the event-based ODMR, we changed the microwave frequency $\omega_{MW}/(2\pi)$ in steps of 1 kHz every 1$\mu s$ from 2836 MHz to 2905 MHz. According to the time resolution of the event camera, we set a duration $T_{event} = 10$ $\mu s$. The light intensity $I$ was defined as the sum of photons emitted by all NV centers in the duration $T_{event}$. We then considered the light intensity difference between two adjacent durations $\Delta I = I_{latter} - I_{former}$, and compared it with a predefined threshold $c_{th} = 1$. If $\Delta I \geq c_{th}$ ($\Delta I \leq -c_{th}$). We then recorded an event 1 (-1). Since the microwave frequency changes by 1MHz every $100T_{event}$ time, we added up the events generated in this frequency interval, then obtained the events number corresponding to the midpoint frequency of this interval, as shown in the bottom panel of Fig. 2A.

**Frame-based ODMR measurement.** The frame-based ODMR was performed following the protocols shown in Supplementary fig. S2. The 532nm laser was kept on throughout the measurement to perform a continuous-wave (CW)-mode quantum sensing. The MW was swept from $f_1$-2836MHz to $f_{70}$-2905MHz with a discrete step of 1MHz. The time duration for one frequency step is $t_{step}$ (different $t_{step}$ from 26ms to 260ms are tried in the measurement) and the total time for a full sweep cycle is T=$t_{step}$*70. During the stepped sweep, the absolute light intensity was recorded by the EMCCD, yielding a series of frames. The frequency sweep and EMCCD detection were synchronized by sequenced external pulses with the same step time. The frequency-tuned light intensities were fitted with the Lorentzian function after which the resonance frequency was extracted. The precision of the sensing was evaluated by calculating the standard deviation of resonance frequency from repeated measurements. Here a binning size of 20 by 20 pixels was used for an improved precision, which means the light intensities stored in 20 by 20 pixels are first summarized before being fitted.

**Event-based ODMR measurement.** The event-based ODMR was performed following the protocols shown in Supplementary fig. S3. Again, the 532nm laser was kept constant throughout the measurement. Triggered with an external pulse, the MW frequency was swept linearly from $f_1$-2836MHz to $f_{70}$-2905MHz with a period T (T changes from 3.5ms to 140ms). During frequency sweeping, the fluorescence change was continuously detected by the event camera and a stream of events was output. As discussed in the working principle in the main text, we used a moving sum method to process those raw events and reconstruct the derivative Lorentzian spectrum. Specifically, a window covering 1MHz MW frequency was used to slide across the full sweeping range, during which all event values generated from this frequency range were summed. The same binning size of 20 by 20 pixels was chosen for processing the raw data as the one used in the frame-based ODMR. Moreover, the measurement mentioned above was repeated for 10 loops and the outputs were stacked to mitigate the influence of noise events. Finally, the processed results were fitted with the derivative Lorentzian function to extract the resonance frequency. The standard deviation of 10 fittings was calculated to describe the sensing precision.

**Calculation of sensing precision and time.** In traditional quantum sensing, the precision is usually defined as sensitivity: $\eta_B = \delta B_{min} * \sqrt{\tau_e} = \frac{\sigma_{f_0}}{\gamma_g} * \sqrt{\tau_e}$ for magnetic field sensing or $\eta_T = \delta T_{min} * \sqrt{\tau_e} = \frac{\sigma_{f_0}}{dD/dT}\sqrt{\tau_e}$ for temperature measurement, where $\sigma_{f_0}$ is the standard deviation of measured resonance frequency $f_0$, $\gamma_g$ is gyromagnetic ratio[12], $dD/dT$ is the thermal susceptibility of the Zero-Field-Splitting energy[74]. Here $\tau_e$ is the exposure time and for traditional-camera-based quantum sensing, $\sigma_{f_0} \sim \frac{1}{\sqrt{\tau_e}}$ if the measurement is shot-noise limited[56]. The event camera, however, only measures the change of photo-current rather than integrating photo-generated charges, so the concept of exposure time does not apply (or is extremely short considering the pixels response can reach μs level) for an event camera. To make a fair comparison, we defined the precision directly as the standard deviation of extracted $f_0$, which can be calculated as $\sigma_{f_0} = \sqrt{\frac{\sum_{i=1}^{N_{repeat}}(f_0^i - \bar{f})^2}{N_{repeat}}}$, where $\bar{f}$ is the mean of $f_0$ measured from different repeated sweeps and $N_{repeat}$ is the repeated measurements. Additionally, for a fair sensing speed comparison, we used the total sensing time τ, i.e. the total measurement time consumed before getting the resonance frequency, to represent imaging speed for all the three methods. In general, the shorter the time τ, the higher the sensing speed. In the event camera measurement, the total time $\tau = T*N_{loop}*2$, where T is the time consumed for one single direction sweeping, $N_{loop}$ is the number of looped sweeps for one measurement, and 2 means forward and backward sweeping. For EMCCD, $\tau = \tau_e + \tau_o$, where $\tau_e$ is the total exposure time for one complete ODMR sweep. Here, $\tau_o$ =1.12s is the data readout and transfer time for the EMCCD we used, and it is this value that limits the further reduction of sensing time cost.

For the works using lock-in cameras[29, 30], the sensing time and precision are transformed from the data provided in the paper. Specifically, for reference 29, they obtained a magnetic sensitivity of $\eta_B = 731nT/\sqrt{Hz}$ using $\tau_e$ =4.8ms averaging time per frame. According to the definition of sensitivity mentioned above, the standard deviation of extracted $f_0$, i.e. sensing precision is $\sigma_{f_0} = \eta_B * \frac{\gamma_g}{\sqrt{\tau_e}} \approx 0.29MHz$. Next the sensing time is estimated to be $\tau_e*N_{in}$=0.34s, where $N_{in}$, i.e. the number of frequency steps swept for one ODMR, is assumed to be the same as our measurement as no specific information is provided in the paper. For reference 30, $\sigma_{f_0} = 0.056MHz$ with 10 repeated acquisitions can be extracted from Fig. 10 in the paper. They used a demodulation frame rate of 3500fps to acquire 500 frames per ODMR measurement, so that the sensing time could be calculated as: 1/3500*500*10=1.43s.

**Dynamic temperature measurement.** Another red laser (MDL-III-637) was added to the initial ODMR system for controlling the temperature of the sample (Fig.4A). As illustrated in the fabrication process, the sample was covered with gold nano-particles which absorb laser power and can be used for local heating of the diamond sample. Due to the high thermal conductivity of the bulk diamond sample, the local heat will transmit across the sample to reach a thermal equilibrium. Hence, the diamond sample temperature can be controlled by modulating the red laser power. The red laser power was modulated by rotating a linear polarizer (LPNIRE100-B) mounted on a motorized rotation stage. For incident power $I_0$ entering the polarizer, the output power is $I_0 \cos^2(\theta)$, where $\theta$ is the angle between the polarization axis of incident laser and polarizer.

For static measurement, the red laser was set to different power levels (from 0 to the maximum of 240mW) by rotating the polarizer to different angles. The resonance frequencies $f_0$ of different pixels (covering around 60*60 pixels, i.e. 16*16 μm² of the ROI) were extracted and transferred to a temperature difference using $\Delta T = (f_0^p - f_0^0)/(\frac{dD}{dT})$, where $f_0^p$ and $f_0^0$ are resonance frequencies measured when the red laser power is p and 0mW (taken as the reference zero temperature which is also used in the dynamic measurement), respectively. The measurement was repeated for 10 times, from which we calculated the temperature precision. For dynamic temperature measurement, the red laser power was continuously changed by rotating the polarizer with the speed ω predetermined by a Python program. During measurement, fixed period P of 14ms and 7ms were tried to sweep the MW frequency and events of 10 looped sweeps were stacked to extract the resonance frequency. It took 0.28s /0.14s (considering forward and backward sweeps) to obtain one temperature distribution. The temporal change of temperature was fitted with function $\Delta T = A_0 cos^2(\omega t + \varphi) + c$, from which we extracted the polarizer's rotation speed ω and the temperature change range $A_0$.

**Acknowledgments**

We thank Prof. Edmund Y. Lam and Dr. Hayden K.H. So for fruitful discussions.
**Funding:** Z.Q.C. acknowledges financial support provided under the HKSAR Government Research Grants Council (RGC) Research Matching Grant Scheme (Grant No. 207300313); the HKSAR Innovation and Technology Fund (ITF) through the Platform Projects of the Innovation and Technology Support Program (Grant No. ITS/293/19FP); HKU Seed Fund; the Guangdong Special Support Project (2019BT02X030); and the Health@InnoHK program of the Innovation and Technology Commission of the HKSAR Government. C.L. acknowledges the HKSAR Government RGC Early Career Scheme (ECS) grant (Grant No. 27210321); NSFC Excellent Young Scientist Fund (Hong Kong and Macau) under Grant 62122005. N.W. acknowledges financial support provided under the HKSAR Government RGC Theme-based Research Scheme (TRS) grant (Grant No. T45-701/22-R). JW acknowledges funding from the EU via the project AMADEUS, the BMBF via the cluster4future QSENS as well as the DFG via the GRK2642 and FOR 2724. YL and ZW acknowledge support from the National Natural Science Foundation of China (Grant No. 12074131) and the Natural Science Foundation of Guangdong Province (Grant No. 2021A1515012030). Y.Z. acknowledges support from the Guangdong Special Support Project (2019BT02X030).
**Author contributions:** Z.Q.C. and C.L. conceived the idea. Z.Y.D., M.G., F. X., K.Z., J.H.Z. performed the experiments and analyzed data under the supervision of Z.Q.C., C.L., and N.W.. Y.Y.L performed the simulations under the supervision of Z.Y.W.. Y.Z. and J.W. discussed the results and commented on the manuscript. Z.Y.D., C.L., and Z.Q.C. wrote the manuscript with input from all authors
**Competing interests:** The authors declare the following competing financial interests: A PCT Patent application has been filed with application No. PCT/CN2022/129840